\documentclass[12pt]{article}

\usepackage{times}
\usepackage{epsfig,amsmath}
\usepackage{subfigure}
\usepackage{graphicx}
\usepackage{color}
\usepackage{epstopdf}
\usepackage{braket}
\usepackage{bm}
\usepackage{threeparttable}
\usepackage{array}
\usepackage{colortbl}
\usepackage{url}
 \usepackage{docmute}

\newenvironment{sciabstract}{%
\begin{quote} \bf}
{\end{quote}}

\newcounter{lastnote}

\title{Photon-Inter-Correlation Optical Communication}
\author{Zeng-Quan Yan,$^{1,2,\ast}$ Cheng-Qiu Hu,$^{1,2,\ast}$ Zhan-Ming Li,$^{1,2}$ Zhong-Yuan Li,$^{3}$\\ Hang Zheng,$^{1,2}$ Xian-Min Jin$^{1,2,\dagger}$\\
 \\
\normalsize{$^1$Center for Integrated Quantum Information Technologies (IQIT), School of Physics}\\
\normalsize{and Astronomy and State Key Laboratory of Advanced Optical Communication Systems}\\
\normalsize{and Networks, Shanghai Jiao Tong University, Shanghai 200240, China.}\\
\normalsize{$^2$CAS Center for Excellence and Synergetic Innovation Center in Quantum Information and}\\
\normalsize{Quantum Physics, University of Science and Technology of China, Hefei, Anhui 230026, China}\\
\normalsize{$^3$Beijing Institute of Astronautical Systems Engineering, }\\
\normalsize{China Academy of Launch Vehicle Technology, Beijing 100076, China}\\
\normalsize{$^\ast$The authors make equal contributions to the work.}\\
\normalsize{$^\dagger$E-mail: xianmin.jin@sjtu.edu.cn}\\
}

\date{}

\begin{document}
\baselineskip24pt

\maketitle

\begin{sciabstract}
The development of modern technology extends human presence beyond cislunar space and onto other planets, which presents an urgent need for high-capacity, long-distance and interplanetary communication. Communication using photons as carriers has a high channel capacity, but the optical diffraction limit in deep space leads to inevitable huge geometric loss, setting an insurmountable transmission distance for existing optical communication technologies. Here, we propose and experimentally demonstrate a photon-inter-correlation optical communication (PICOC) against an ultra-high channel loss. We treat light as a stream of photons, and retrieve the additional information of internal correlation and photon statistics globally from extremely weak pulse sequences. We successfully manage to build high-fidelity communication channel with a loss up to 160dB by separating a single-photon signal embedded in a noise ten times higher. With only commercially available telescopes, PICOC allows establishment of communication links from Mars to Earth communication using a milliwatt laser, and from the edge of the solar system to Earth using a few watts laser.
\end{sciabstract}

After a 42-year trek, Voyager 1 has finally crossed the heliosphere, becoming the first artifact in human history to enter interstellar space. Voyager 1 was equipped with advanced radio frequency communication terminals at the time, but due to the low radio frequency channel capacity and large divergence angle, it is currently only able to establish a communication link with Earth at $\sim$100 bit/s\cite{Voyager_communication}. Meanwhile, the scope of human presence is gradually expanding to other planets in the solar system, which presents an urgent need for high-capacity, long-distance interplanetary communication with reasonable system size.

Laser communication, with directivity and channel capacity superior to radio frequency communication, is becoming the alternative solution of the next generation of interstellar communication \cite{Coherent_detection_Kazuro_2006,Coherent_communication_Satoshi_2008}. However, the diffraction limit of optical systems leads to the inevitable beam divergence associated with geometric loss, which restricts the practical communication distance to a million kilometers \cite{NASA_Mars_FSO,NFIRE_TerraSAR-X,NFIRE_Ground,CWB:MCLCD,LLCD1,LLCD2}. Taking the Mars-to-Earth optical communication as an example, the spot diameter would exceed 600 kilometers using the transmitting telescope with a 10$\mu rad$ divergence angle. Meanwhile, the aperture of the receiving optical telescope is limited to a few meters, which leads to a huge geometric loss as high as 110dB. Considering system loss, beam wandering, and reasonable laser power, the received signal intensity will fade far below one photon per pulse.

Furthermore, noise, no matter from environment or detector, is another critical factor and becomes dominant to the signal-to-noise ratio (SNR) when the signal drop to the noise level. Eventually, the error rate goes too high to establish a reliable communication link with the SNR below 0dB. Pioneering works have introduced the single-photon detector (SPD) into optical communication to improve system sensitivity and loss tolerance\cite{Quantum_receiver_2013,Quantum_receiver_2015,Quantum_receiver_SNDA_2020,SPAD_communication_2014,Underwater_SPAD_2016}, straightforwardly taking advantage of the detector's single-photon response (see Fig.1).

Here, we treat light as a stream of photons in a picture of quantum optics. Each photon in the stream has many degrees of freedom including time, polarization, frequency, phase and orbital angular momentum, and much more implicit information can be extracted from many more photons globally. We propose and experimentally demonstrate an photon-inter-correlation optical communication (PICOC) exploiting time information globally shared in millions of pulses. The retrieved global information shared in the survived photons enables optical communication in the extreme condition of high loss, negative SNR, and frequent channel disconnection, which is intractable with conventional schemes. Moreover, we reveal the time evolution of photon statistics with high resolution, realizing the precise and optimal manipulation of bit error rate (BER) during the communication.

\subsection*{Scheme of PICOC}
\textcolor{black}{
PICOC allows us to retrieve the time frame from the sparse single-photon stream and synchronize signals under ultra-high loss scenarios. Time synchronization is the prerequisite of all kinds of communication. However, the phase drift between remote clocks is inevitably affected by temperature fluctuation, voltage disturbance and the doppler effect, which restrict the establishment of communication links\cite{Optical_challenges,Satellite_optical_com_doppler}. Even using atomic clocks, the phase drift induced by the time dilation becomes the key barrier of effective communication, especially between fast-moving satellites\cite{GPS_time_dilation,Space_clock,Verify_time_dilation}.
}

Optical communication based on photodetectors requires hundreds of thousands of signal photons per pulse to establish reliable communication links\cite{Underwater_2.7G,Underwater_100m_500M}. High intensity of received signals reduces the difficulty of synchronization but at the same time immensely limits the maximum tolerable channel loss. With the application of SPDs, the system can tolerate more channel loss (see Fig.1). However, they still require dozens of photons per pulse to perform the photon-counting pulse synchronization scheme\cite{SCPPM,PCPS}.

In our experiment, only about $10^{-5}$ photon per pulse survived, the received signal intensity is as weak as that of a quantum key distribution\cite{XiongfengMa:QKD,Hoi_Kwong_Lo:QKD}. Synchronization methods of prior arts are incapable of retrieving the time frame. One solution is to use an assistant timing pulse, for example, a high-power low-repetition pulse laser\cite{QKD_122km_fiber,QKD_freespace_100km,QKD_hotbolloon,QKD_satellite,QKD_55m_underwater,QKD_30m_underwater} or a GPS pulse-per-second signal\cite{QKD_daylight,QKD_daylight_chip}.
\textcolor{black}{
Here, instead of introducing an assistant signal, our PICOC scheme provides a passive way to extract time frames shared in the sparse single-photon stream. Stage1: We record the arrival time of each photon using the receiver's time reference. By dividing time into many 2ms length time slots and searching the coincidence delay in each slot, we obtain the time frame shared among several photons (see Fig.2b). We can observe the coincidence delay oscillating with time at this stage, representing the phase drift between remote clocks. Stage2: We perform the Fourier transformation to analyze the overall coincidence delay. The global analysis feature of Fourier transformation allows us to correlate many more photons and retrieve the time frame from various combinations of photon groups (see Fig.2c). The first peak in Fig.2c represents that the linear phase drift during the 300s measurement is around 7415.77 pulse periods. We assume that the linear phase drift is uniformly distributed in all pulses. By further retrieving the linear phase drift of each pulse, we can calibrate photons' arrival time and achieve coarse time synchronization. Stage3: More generally, we also consider the non-linear phase drift between remote clocks after performing the linear phase drift calibration. By searching the coincidence delay of each second, we can characterize the non-linear phase drift between remote clocks (see Fig.2d). Finally, we draw all photons into a pulse period and retrieve the pulse waveform (see Fig.2e). The time synchronization is well established compared with the result without our scheme.
}

\textcolor{black}{
PICOC synchronization procedure is a passive synchronization method based on the arrival time of photons. We use photons as carriers for both message and time information. By correlating the internal time information shared among survived photons globally, we manage to establish the high precision time synchronization from the sparse single-photon stream with only $10^{-5}$ photon per pulse is received.
}

\textcolor{black}{
PICOC also allows us to precisely manipulate the communication performance against ultra-high loss. The basic idea is to use the time evolution of photon statistics to separate signal and noise. Here, we use empty pulses and non-zero light pulses to represent 0-bits and 1-bits in the message. By correlating photons from the extremely weak repetitive sequence, we can extract cyclically redundant photons for each bit. Then, we count the correlated photon number for each bit and obtain the photon statistics (see Fig.2g and Fig.2h). By analyzing data for a longer time, we can correlate many more photons and obtain a different distribution pattern. In Fig2.g and Fig.2h, we show the theoretical and experimental time evolution of photon statistics. In the beginning, the distribution of 0-bits (the left peak) and 1-bits (the right peak) overlap each other, making it hard to set a threshold and tell signal from noise. However, the time-evolving behavior of photon statistics shows the ability to gradually separate these distributions, giving the possibility to retrieve the bit sequence with a lower BER. In other words, the longer we measure, the more photons we correlate, and the lower BER we obtain (see Fig.2i). Eventually, photon statistics of 1-bits and 0-bits separate far enough to establish a reliable communication link, as long as we manage to synchronize signals from these extremely sparse single-photon streams.
}

\textcolor{black}{
Light decays into weak coherent-state signals when the attenuation is huge. The received photon number in each signal pulse follows a Poisson distribution\cite{XiongfengMa:QKD,Hoi_Kwong_Lo:QKD}. Meanwhile, random noise from the background forms a Poisson process. The noise photon number in each pulse is also a random variable with a Poisson distribution\cite{poisson_process}. Bit error rate (BER) of PICOC can be expressed as
}
    \begin{equation}
        \left\{
        \begin{aligned}
            BER(t_{ac}) &=a_{0}\left[1-\frac{\Gamma\left(N_{s}, \mu_{0} t_{ac}\right)}{\Gamma\left(N_{s}\right)}\right]+a_{1} \frac{\Gamma\left(N_{s}, \mu_{1} t_{ac}\right)}{\Gamma\left(N_{s}\right)} \\
            N_{s}(t_{ac}) &=\left\lceil\frac{\ln \left(a_{0} / a_{1}\right)+t_{ac}\left(\mu_{1}-\mu_{0}\right)}{\ln \left(\mu_{1} / \mu_{0}\right)}\right\rceil
        \end{aligned}
        \right.
    \end{equation}
where $\mu_{0}$ and $\mu_{1}$ indicate the average correlated photon number of 0-bits and 1-bits per second respectively. $a_{0}$ and $a_{1}$ represent the percentage of 0-bits and 1-bits. The accumulation time is expressed as $t_{ac}$. \textcolor{black}{The gamma function $\Gamma(x)$ and the upper incomplete gamma function $\Gamma(s,x)$ can be expressed as $\Gamma(x)=\int_{0}^{+\infty} t^{x-1} e^{-t} \mathrm{d} t$ and $\Gamma(s, x)=\int_{x}^{+\infty} t^{s-1} e^{-t} \mathrm{d} t$ respectively. The photon number threshold $N_{s}$ retrieves the bit sequence with the minimum error probability. When the correlated photon number of a bit is higher than $N_{s}$, we consider it as a 1-bit and vice versa.}

Apparently, we can precisely manipulate BER by setting the value of accumulation time (see Fig.2f). However, the gamma function $\Gamma$ and the step function $N_{s}(t_{ac})$ in Eq.1 increase the difficulty in solving the value of $t_{ac}$. Here, we give an approximate expression of BER, which can be numerically solved in real time.
    \begin{equation}
        \ln \left[BER(t_{ac})\right]=G_{cc} \left( t_{ac} +\varphi_{N_{s}} \right) +p+A_{p}\left|\sin \left[\omega\left(t_{a c}-\varphi_{N_{s}}\right)\right]\right|
    \end{equation}

\textcolor{black}{
The relationship between BER and $t_{ac}$ consists of two terms, the linear term $\left[ G_{cc} \left( t_{ac} +\varphi_{N_{s}} \right) +p \right]$ and the oscillating term $A_{p}\left|\sin \left[\omega\left(t_{a c}-\varphi_{N_{s}}\right)\right]\right|$. The linear term perceives light as a continuous field, obtaining a smooth BER variation with time. Dominant factors such as coarse correlation gain $G_{cc}$ and initial phase $p$ determine the overall time evolution tendency of BER. The oscillating term treats light as a stream of photons from the picture of quantum optics. By analyzing the step function in Eq.1, we can calculate the single-photon related oscillation frequency $\omega$ and phase $\varphi_{N_{s}}$ in our system (see Supplement Material for details). $A_{p}$ is the amplitude of the oscillating term, which is also a single-photon related parameter. With the increase of noise photons, BER contributed by the oscillating term tends to decrease. Meanwhile, the time evolution of BER degenerates to the classical form, where light is only treated as a continuous field (see Fig.2f).}

\textcolor{black}{
After achieving the precise manipulation of BER, we should further consider the time efficiency of the PICOC system. The derivative of $\ln \left[BER\left(t_{ac}\right)\right]$ with respect to accumulation time $t_{ac}$ characterizes the correlation gain $G_{c}$ enhanced by time. When we treat light as a stream of photons, we can reveal the fine periodic structure of the correlation gain $G_{c}$ over time. The sensitivity of BER to time depends on the value of $G_{c}$. The higher the value of $G_{c}$, the faster the BER drops with time. Once the value of $G_{c}$ drops to zero, time becomes irrelevant with BER. Therefore, we can optimize our system's time efficiency by manipulating the value of $G_{c}$. However, directly calculate the correlation gain from Eq.1 is difficult. Here, we derive the approximate expression of correlation gain from Eq.2.
}
    \begin{equation}
        \begin{aligned}
        G_{c} &=-\frac{\mathrm{d} \ln \left[BER\left(t_{ac}\right)\right]}{\mathrm{d} t_{a c}} =-G_{cc}-A_{p} \omega \frac{\sin \left[2 \omega\left(t_{a c}-\varphi_{N_{s}}\right)\right]}{2 \vert\sin \left[\omega\left(t_{a c}-\varphi_{N_{s}}\right)\right]\vert}
        \end{aligned}
    \end{equation}

\subsection*{Experiment and results}
\textcolor{black}{
In the transmitting terminal, a green laser diode is driven to 50MHz (5ns pulse width) using return-to-zero on-off keying modulation. By transmitting the 125k-bit message cyclically, we distribute redundant pulses in time uniformly. The global encoding method guarantees that the manipulation of communication performance in our system is only related to measured time. The decoding set-up consists of a coupler module, a SPD, and three process modules (see Fig.2a). In the first process module, we record the arrival time of photons with a 64ps time resolution. By performing the PICOC time synchronization procedure, the sparse single-photon stream is synchronized within 500ps, close to SPD's pulse duration (see Fig.2e). Meanwhile, we can also retrieve the pulse position $s_{i}(i=1,2,3,...)$ of each photon. In the second process module, we further correlate the photon number of each bit. The corresponding bit position of a photon can be expressed as $(s_{i} \mod l)+1$, where $l$ represents the length of the transmitted message. By correlating all photons, we can obtain the time evolution of photon statistics and estimate the BER performance (see Fig.2f). In the last process module, we choose the optimal photon number threshold $N_{s}$ and retrieve the message with the minimum BER (see Fig.2i).
}

The 520nm laser diode is a cost-effective source, and most importantly the thumb-sized laser can well carried by drones or satellites for optical communication with low power dissipation. Fig.3 illustrates BER as a function of accumulation time under various channel loss and noise level. We build the high-fidelity communication link with a loss up to 124dB by only using a 0.552mW laser, which is very encouraging because PICOC equipped with such a small-power laser has already been able to reach a record against loss (see Fig.3c)\cite{High_loss}. Apparently, PICOC equipped with high-power laser can further extend the capacity of loss tolerance in a straightforward way. By employing a commercially available 2W solid-state laser at 532nm, we show that we can achieve optical communication with a channel loss up to 160dB (see Fig.3d).

As is shown in Fig.3, we can always push BER down below the FEC threshold by correlating more photons. Noise here plays a similar role as loss for decreasing SNR, which can be overcome by correlating more photons. Our PICOC can give a complete evolution of BER and the continuous dynamics in terms of loss, noise and time.
\textcolor{black}{
In Fig.3e, we show the correlation gain $G_{c}$ from the PICOC experiment against 160dB channel loss. The analytical and experimental curves are in agreement with the theoretical curve. Every oscillation cycle of $G_{c}$ characterizes a value switch of optimal photon number threshold $N_{s}$. Generally, once the value of $N_{s}$ switches from one to another, $G_{c}$ reaches the high-gain point. However, due to the short accumulation time, the integral value of $G_{c}$ is still in the idle region and the BER is just beginning to decrease. When the optimal threshold is close to the next round switch, the value of $G_{c}$ is extremely small and the BER barely drops with time. Therefore, we should avoid these idle regions to achieve optimal time efficiency by simply setting the value of $G_{c}$ on the mid-gain point.
}

\textcolor{black}{
In our system, the data rate and BER are no longer determined by the system's prior parameters. PICOC can characterize and predict the channel performance during the communication, and realize the precise manipulation of BER on demand.}
By setting BER to the FEC threshold, we can solve the Eq.2 and obtain the required accumulation time, which define the optical communication data rate. \textcolor{black}{In our system, the data rate ranges from 40 bits per second (bps) to 42kbps (see Fig.4).} As noted in Fig.4, the 0dB SNR boundary separates the map into two parts. On the right part, there are more signal photons than noise photons, representing the positive SNR environment. In contrast, on the left parts, PICOC can retrieve unknown signals completely embedded in noise and demonstrates high-fidelity communication channel with a negative SNR down to -10.35dB. The strong robustness of PICOC against extremely low SNR is essential to interstellar optical communication, since the channel loss between the planets would quadratically reduce signal photons while noisy environment is inevitable especially for the round-the-clock need of communication.

We estimate the achievable communication range that our PICOC equipped with currently available devices can reach, as is shown in Fig.5. We take the transmitting telescope of the Micius satellite as the reference\cite{QKD_satellite,Quantum_teleportation_satellite,Quantum_network_satellite}. The transmitting aperture and beam divergence are 300mm and 10$\mu rad$. Meanwhile, the receiving station equipped with a 1200-mm-aperture telescope is conceived. \textcolor{black}{The equivalent channel loss from Mars to Earth and the edge of the solar system to Earth is around 113dB and 160dB, including the 3dB loss caused by atmospheric absorption and turbulence (see Supplemental Material for details)}. Our experiments successfully demonstrate optical communication against the above channel losses using a laser diode and a high peak power laser. Noted that the accumulation time shows the non-exponential growth with distance, representing the establishment of communication links in reasonable time consumption.

\subsection*{Conclusion and discussion}
In summary, we propose an optical communication system harnessing the single-photon feature and internal correlation of light. We experimentally demonstrate its capability against \textcolor{black}{frequent channel disconnection (see Supplemental Material for details)}, ultra-high channel loss and noisy environments. By retrieving the implicit internal information and photon statistics globally from survived photons, we reveal the time evolution of photon statistics with high resolution and realize the precise manipulation of the system performance during the communication. \textcolor{black}{
Compared with achievable technology encouraged by NASA\cite{NASA_Mars_FSO,DSOC_predict,DSOC_trans} and existing technology equipped on Voyager 1\cite{Voyager_communication}, our scheme inspires interstellar optical communication with reasonable telescope aperture and low power dissipation (see Table 1).}

\begin{table}[ht]
    \arrayrulecolor{black}
        \centering
        \begin{threeparttable}
        \caption{\textcolor{black}{\textbf{Demands of different interstellar communication systems.}}}
        \begin{tabular}{rrrrrrr}

            \hline
            {\color{black}System}&{\color{black}Carrier}&{\color{black}Transmitting}&{\color{black}Receiving}&{\color{black}Data rate}&{\color{black}Power}&{\color{black}Distance}\\
             & &{\color{black}aperture}&{\color{black}aperture}& &\\
            \hline
            {\color{black}PICOC}&{\color{black}Photon}&{\color{black}30cm}&{\color{black}1.2m}&{\color{black}$\sim$100bps}&{\color{black}2W}&{\color{black}Termination shock-Earth}\\
            {\color{black}Voyager 1\cite{Voyager_communication}}&{\color{black}Radio wave}&{\color{black}366cm}&{\color{black}34m}&{\color{black}$\sim$100bps}&{\color{black}23W}&{\color{black}Termination shock-Earth}\\
            {\color{black}DSOC\cite{DSOC_predict}}&{\color{black}Photon}&{\color{black}22cm}&{\color{black}12m}&{\color{black}$\sim$10kbps}&{\color{black}4W}&{\color{black}Saturn-Earth}\\
            \hline
        \end{tabular}
        \begin{tablenotes}
            \footnotesize
            \item[]\textcolor{black}{The termination shock is one boundary of the Solar System. Deep space optical communication (DSOC) is an alternative communication scheme for NASA's future outer planet missions.}
        \end{tablenotes}
        \end{threeparttable}
\end{table}

The potential applications include various high-loss scenarios such as underwater wireless optical communication over one kilometer, air-to-sea long-distance time and frequency transmission, and interplanetary optical broadcast. The extraction of time frames from weak signals of quantum key distribution is also achievable. Besides, it would also be interesting to perform PICOC in single-photon side-channel attacks. By retrieving information from a few scattered photons, we may bring powerful eavesdroppers into real life\cite{QKD_hack}.

\subsection*{Funding}
National Key R\&D Program of China (2017YFA0303700); National Natural Science Foundation of China (NSFC) (61734005, 11761141014, 11690033); Science and Technology Commission of Shanghai Municipality (STCSM) (15QA1402200, 16JC1400405, 17JC1400403); Shanghai Municipal Education Commission (SMEC)(16SG09, 2017-01-07-00-02-E00049); Shanghai Municipal Science and Technology Major Project (2019SHZDZX01).

\subsection*{Acknowledgments}
X.-M.J. acknowledges support from the National Young 1000 Talents Plan.

\clearpage


\clearpage

\begin{figure*}[htbp]
    \centering
    \includegraphics[width=1.0\linewidth]{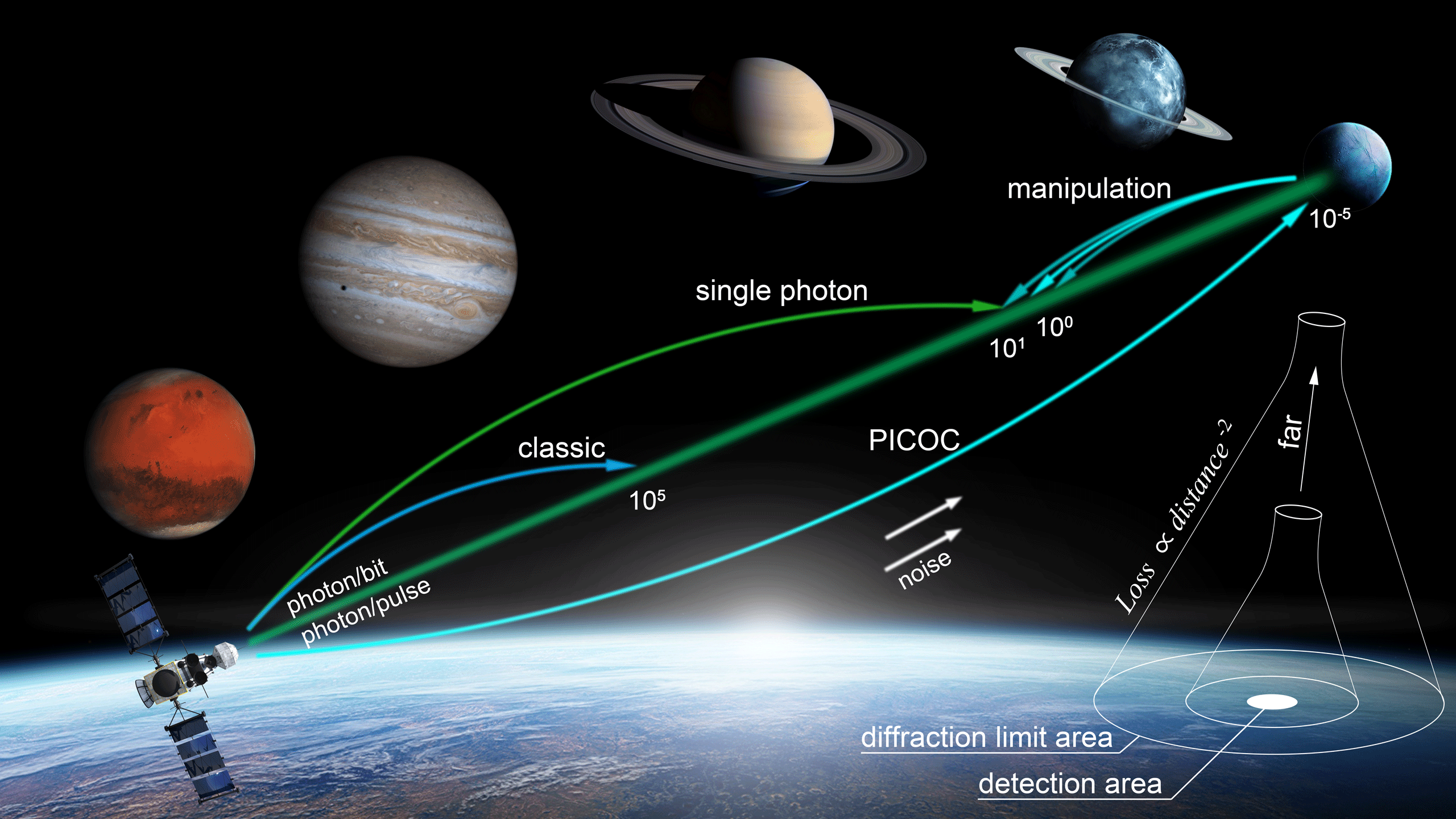}
    \caption{\textbf{Benchmark of achievable capabilities of different communication schemes.}
    Classical optical communication using photodetectors requires hundreds of thousands of photons per pulse to establish reliable links, which limits the maximum transmission channel loss. Optical communication based on single-photon detectors can be much more sensitive to signal but still requires dozens of photons per pulse to build reliable links. In contrast, we treat light as a stream of photons and retrieve the global information shared in very limited survived photons. We can establish the optical communication link with the received signal intensity of $10^{-5}$ photon per pulse. Meanwhile, the desired bit error rate (BER) can be precisely manipulated owing to the global correlation feature of PICOC.
    }
    \label{f1}
\end{figure*}

\clearpage

\begin{figure*}[htbp]
    \centering
    \includegraphics[width=1.0\linewidth]{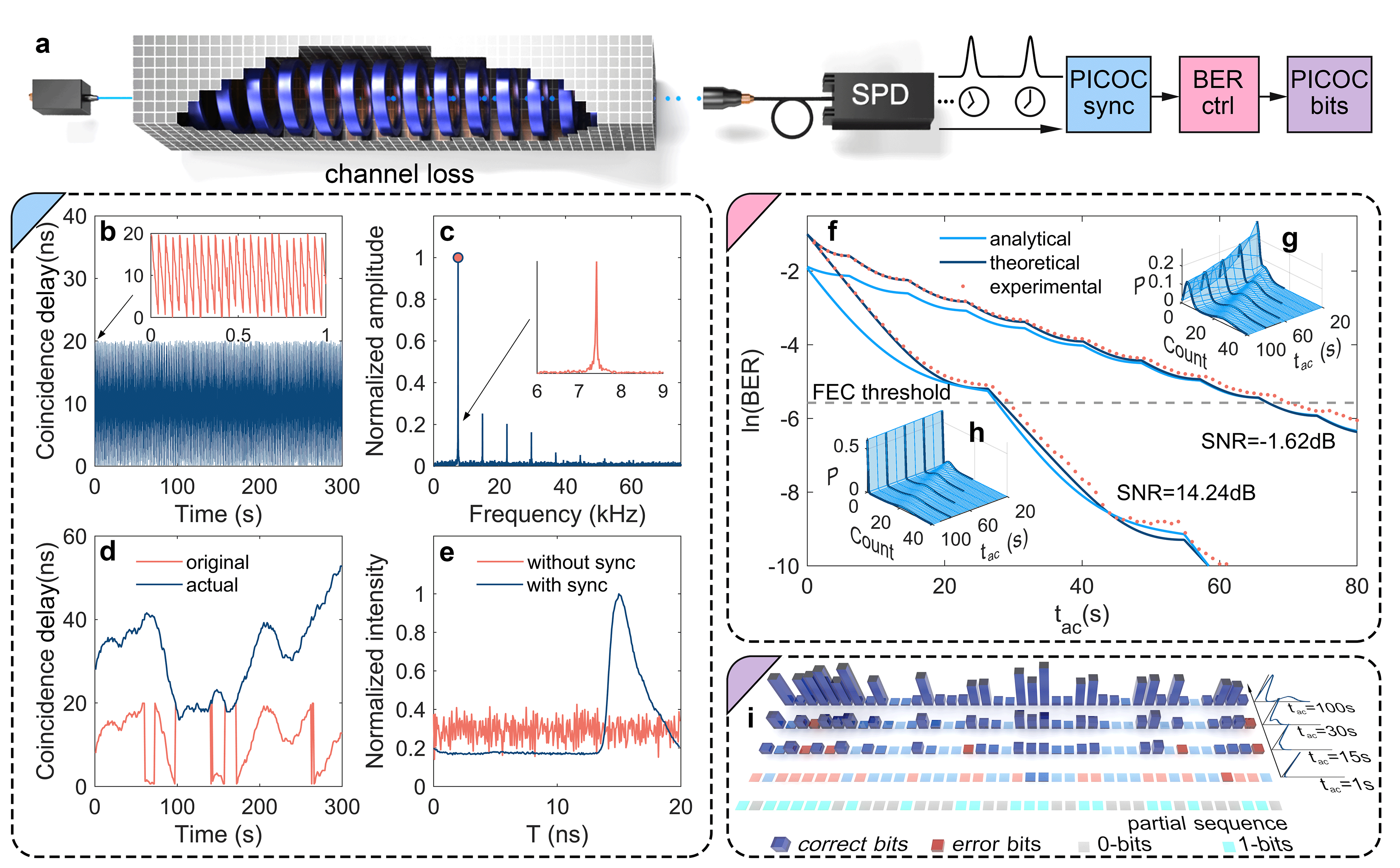}
    \caption{\textcolor{black}{
    \textbf{Establishing reliable communication links using PICOC.}
    \textbf{ a}, Schematic of the experimental setup.
    \textbf{ b}, Retrieving time frames from the sparse single-photon stream. We show the experimental results of 107dB channel loss and -1.62dB signal-to-noise ratio (SNR). Stage1: By searching the coincidence delay of each time slot (2ms duration), we can retrieve the correlation shared among several photons.
    \textbf{ c}, Stage2: We perform a Fourier transformation to analyze the overall coincidence delay. The Fourier transformation can retrieve time frames from various combinations of photon groups, which allows us to correlate many more photons and find the linear phase drift between remote clocks globally. Then we can relocate the arrival time of photons and obtain the coarse time synchronization.
    \textbf{ d}, Stage3: Building on the previous step, we further correlate photons in each second and observe the non-linear phase drift between remote clocks. The fine time synchronization is then achieved by analyzing the time information shared among the sparse single-photon stream.
    \textbf{ e}, The retrieved waveform of the pulse with and without the photon-inter-correlation synchronization procedure.
    \textbf{ f}, Theoretical and experimental BER as a function of the accumulation time with different SNR conditions.
    \textbf{ g}, Time evolution of the photon statistics. We set the channel loss up to 107dB and SNR down to -1.62dB. The surface in the figure represents the expected photon statistics, and the thick solid curve shows the experimental results.
    \textbf{ h}, Time evolution of the photon statistics with 107dB channel loss and 14.24dB SNR.
    \textbf{ i}, Retrieving the bit sequence from the sparse single-photon stream. Histograms show the correlated photon number of bits, together with the photon statistics of the sequence. The longer we measure, the more photons we manage to correlate, and the lower BER we obtain. FEC: forward error correction.
    }}
    \label{f2}
\end{figure*}

\clearpage

\begin{figure*}[htbp]
    \centering
    \includegraphics[width=1.0\linewidth]{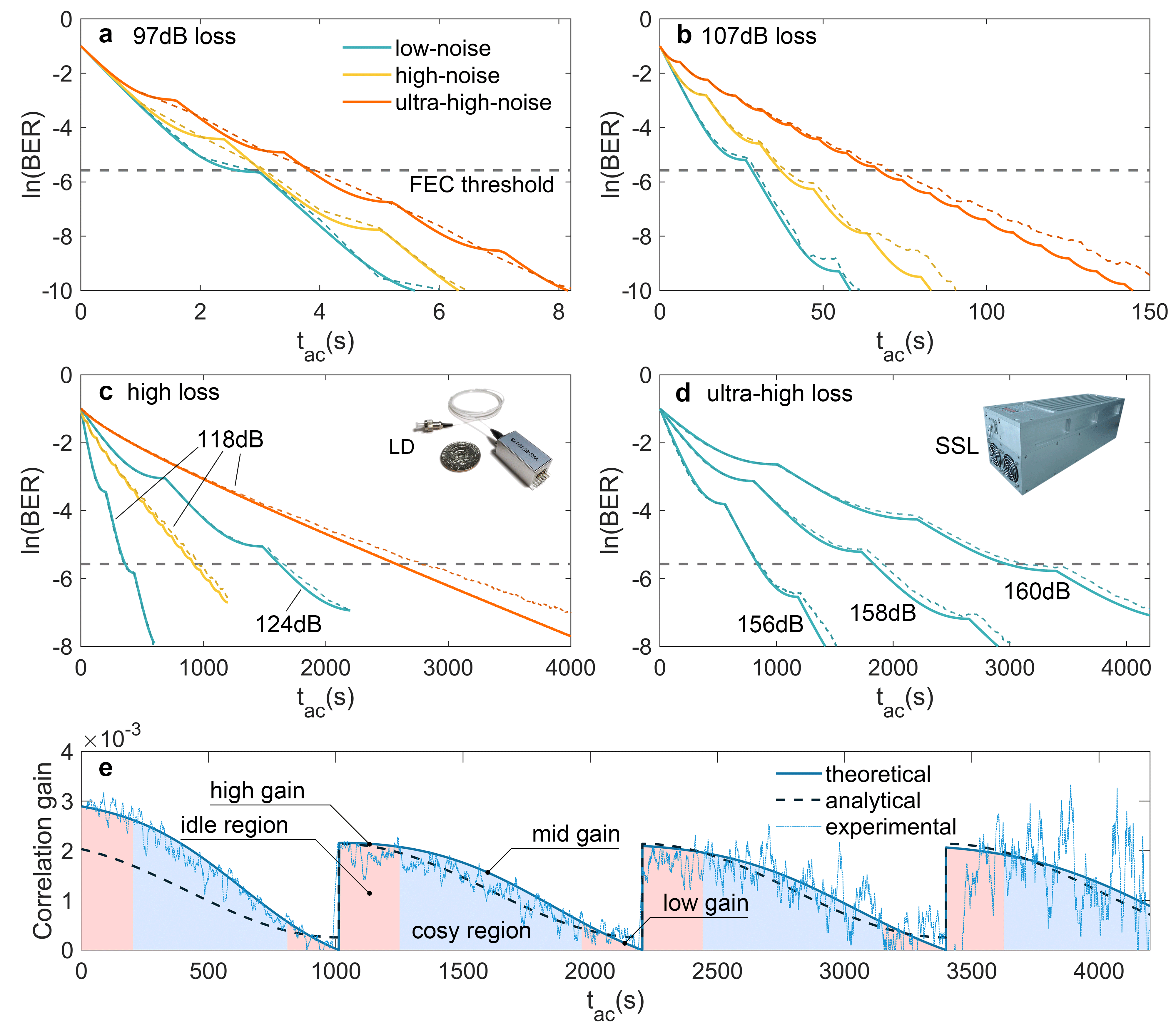}
    \caption{\textcolor{black}{
    \textbf{Experimental BER and correlation gain manipulation enabled by PICOC.}
    \textbf{ a}, Measured BER versus accumulation time with laser diode (LD) and channel loss of 97dB. The solid and dotted curves represent the theoretical and experimental results respectively. The orange, yellow and turquoise curves represent low-noise, high-noise and ultra-high-noise environments.
    \textbf{ b}, Measured BER versus accumulation time with LD and channel loss of 107dB.
    \textbf{ c}, Measured BER versus accumulation time with LD and channel loss of 118dB and 124dB.
    \textbf{ d}, Data obtained with high-power solid-state laser (SSL) against 156dB, 158dB and 160dB channel loss and a low noisy environment.
    \textbf{ e}, Theoretical, analytical and experimental correlation gain as a function of the accumulation time. Data obtained with high-power SSL against 160dB channel loss and a low noisy environment. Low-noise: noise level of 70 counts per second(cps); high-noise: noise level of 2kcps; ultra-high-noise: noise level of 15kcps; LD output power: 0.552mW; SSL output power: 2W.
    }}
    \label{f3}
\end{figure*}

\clearpage

\begin{figure*}[ht]
    \centering
    \includegraphics[width=1.0\linewidth]{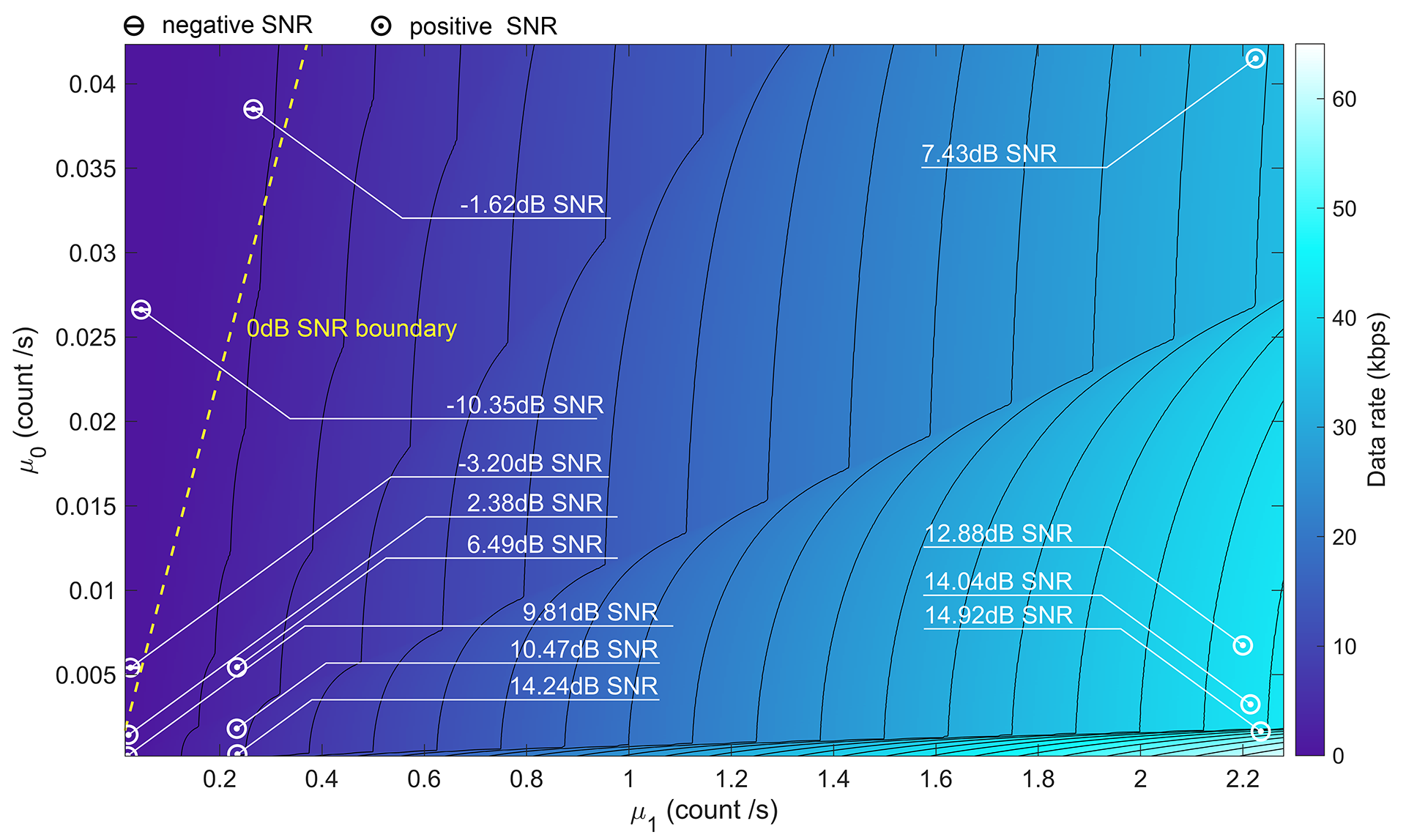}
    \caption{\textbf{Experimental results of data rate with PICOC.}
    The data rate under various noisy environment and channel loss. The value of $\mu_{0}$ and $\mu_{1}$ characterize the noise environment and channel losses respectively. The yellow dashed line divide the map into two parts, where the left part represent negative SNR.
    }
    \label{f4}
\end{figure*}

\clearpage

\begin{figure*}[ht]
    \centering
    \includegraphics[width=1.0\linewidth]{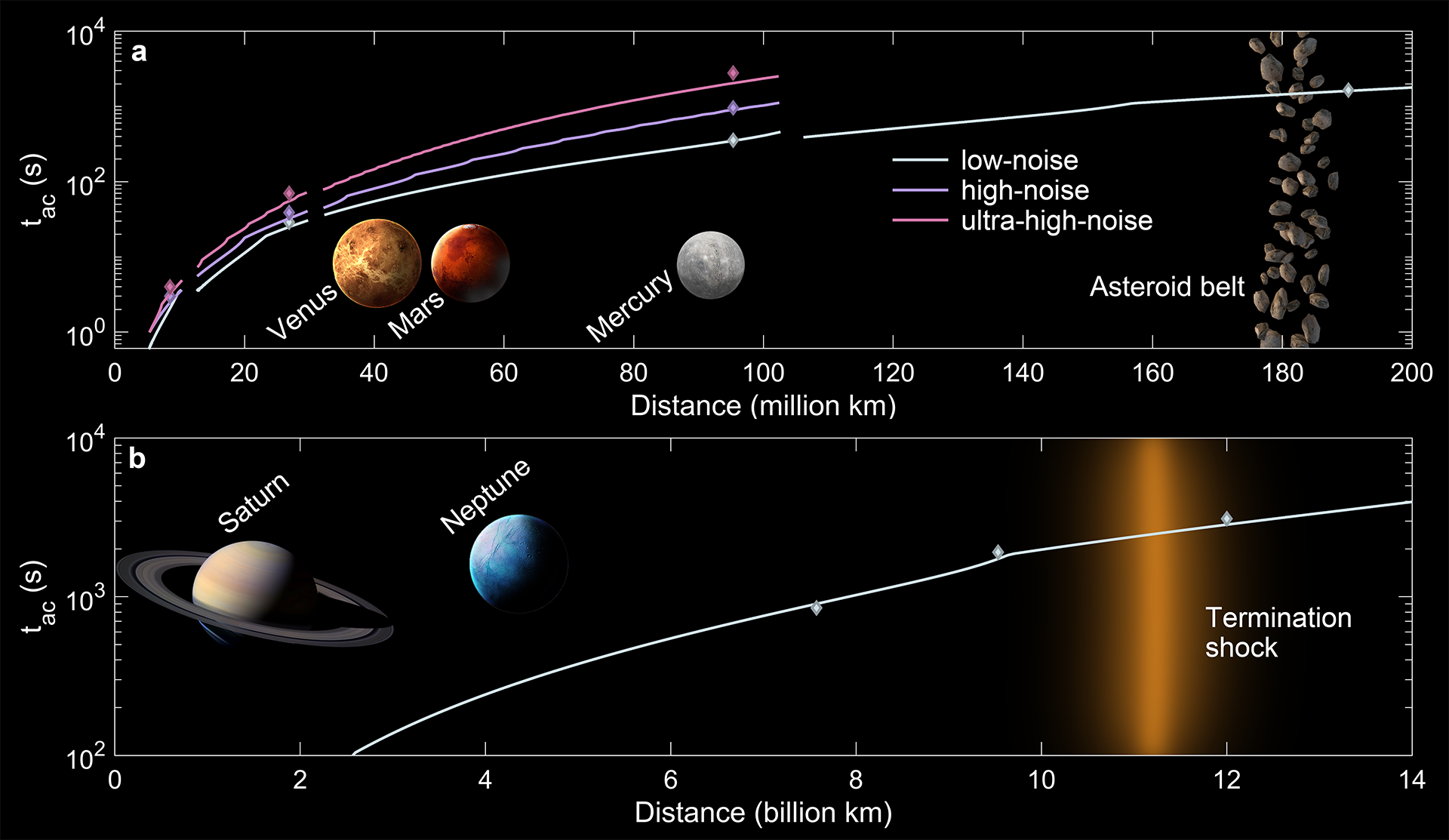}
    \caption{\textbf{Optimal accumulation time for interplanetary communication with PICOC.}
    \textbf{ a}, Accumulation time with the equivalent transmission distance variation. We choose data of 0.552mW laser diode with several different noise level. Low noise, high noise and ultra-high noise indicate the noise level of 70cps, 2kcps and 15kcps.  The solid lines and dots represent the theoretical and experimental results respectively. By mapping the corresponding channel loss to interplanetary communication distance, we obtain the relationship between communication distance and accumulation time.
    \textbf{ b}, Experimental results of demonstrating optical communication with a 2W solid-state laser, marked at the high-loss conditions being equivalent from the edge of the solar system to Earth.
    }
    \label{f5}
\end{figure*}

\clearpage

\appendix

\section*{\centering{Supplemental Material}}
\author{\centerline{Zeng-Quan Yan,$^{1,2,\ast}$ Cheng-Qiu Hu,$^{1,2,\ast}$ Zhan-Ming Li,$^{1,2}$ Zhong-Yuan Li,$^{3}$}
\centerline{Hang Zheng,$^{1,2}$ Xian-Min Jin$^{1,2,\dagger}$}
\centerline{\textcolor{white}{space}}

\centerline{$^1$Center for Integrated Quantum Information Technologies (IQIT), School of Physics}
\centerline{and Astronomy and State Key Laboratory of Advanced Optical Communication Systems}
\centerline{and Networks, Shanghai Jiao Tong University, Shanghai 200240, China.}
\centerline{$^2$CAS Center for Excellence and Synergetic Innovation Center in Quantum Information and}
\centerline{Quantum Physics, University of Science and Technology of China, Hefei, Anhui 230026, China}
\centerline{$^3$Beijing Institute of Astronautical Systems Engineering, }
\centerline{China Academy of Launch Vehicle Technology, Beijing 100076, China}
\centerline{$^\ast$The authors make equal contributions to the work.}
\centerline{$^\dagger$E-mail: xianmin.jin@sjtu.edu.cn}
}

\centerline{\textcolor{white}{space}}
\centerline{\textcolor{white}{space}}

\renewcommand\thefigure{S\arabic{figure}}
\setcounter{figure}{0}
\renewcommand{\theequation}{S\arabic{equation}}
\setcounter{equation}{0}

\subsection*{1. Establishment of PICOC links under frequent channel disconnection}
To simulate the channel disconnection, we use a black paperboard to intercept the optical link. From the 38th to the 39th second, about 50 million signal pulses are completely lost (see Fig.S1a). The bit error rate (BER) goes too high to establish a reliable communication link for conventional technology. However, the global correlation feature of PICOC allows us to retrieve the sequence between remote pulses, even they are millions of pulses apart. By further retrieving redundant photons distributed in billions of pulses, we manage to demonstrate the communication against frequent channel disconnection (see Fig.S1b).

\begin{figure*}[htbp]
    \centering
    \includegraphics[width=1.0\linewidth]{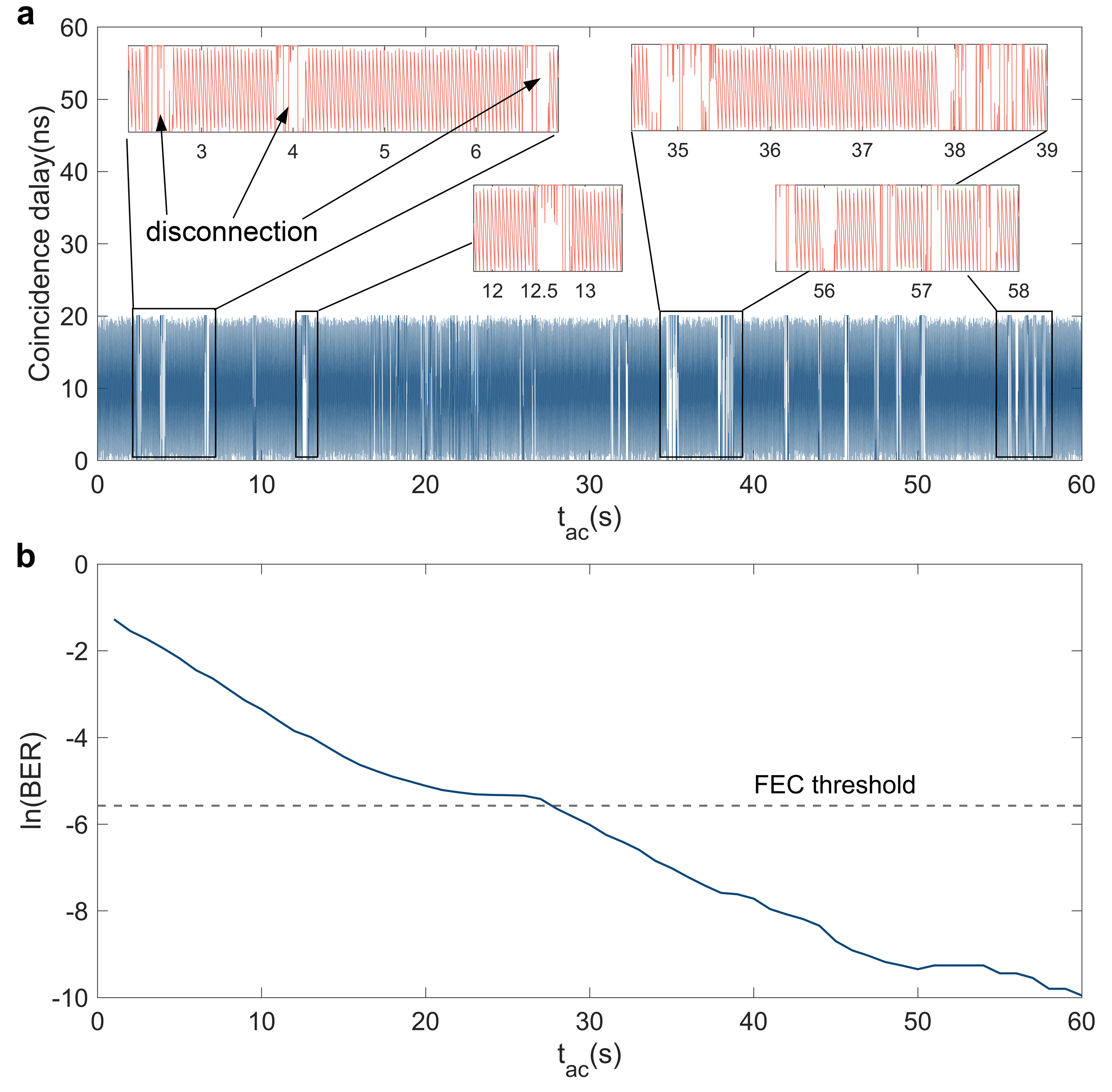}
    \caption{\textbf{Establishing reliable optical communication links against frequent channel disconnection.}
    \textbf{ a}, Retrieving time frames from the sparse single-photon stream under frequent channel disconnection scenario. We show experimental results under 107dB channel loss.
    \textbf{ b}, Time evolution of the bit error rate under frequent channel disconnection scenario.
    }
    \label{f1}
\end{figure*}

\subsection*{2. Outdoor verification on the properties of the measurement process}
We assume that the signal and noise photon number in each pulse is a random variable with a Poisson distribution. To justify our assumption, we demonstrate an outdoor experimental verification on the properties of the measurement process. We use a commercially available drone to carry the PICOC transmitter (see Fig.S2). By increasing the beam divergence to a 90-degree angle, we manage to simulate the huge light spot from planet to planet.

\begin{figure*}[htbp]
    \centering
    \includegraphics[width=0.7\linewidth]{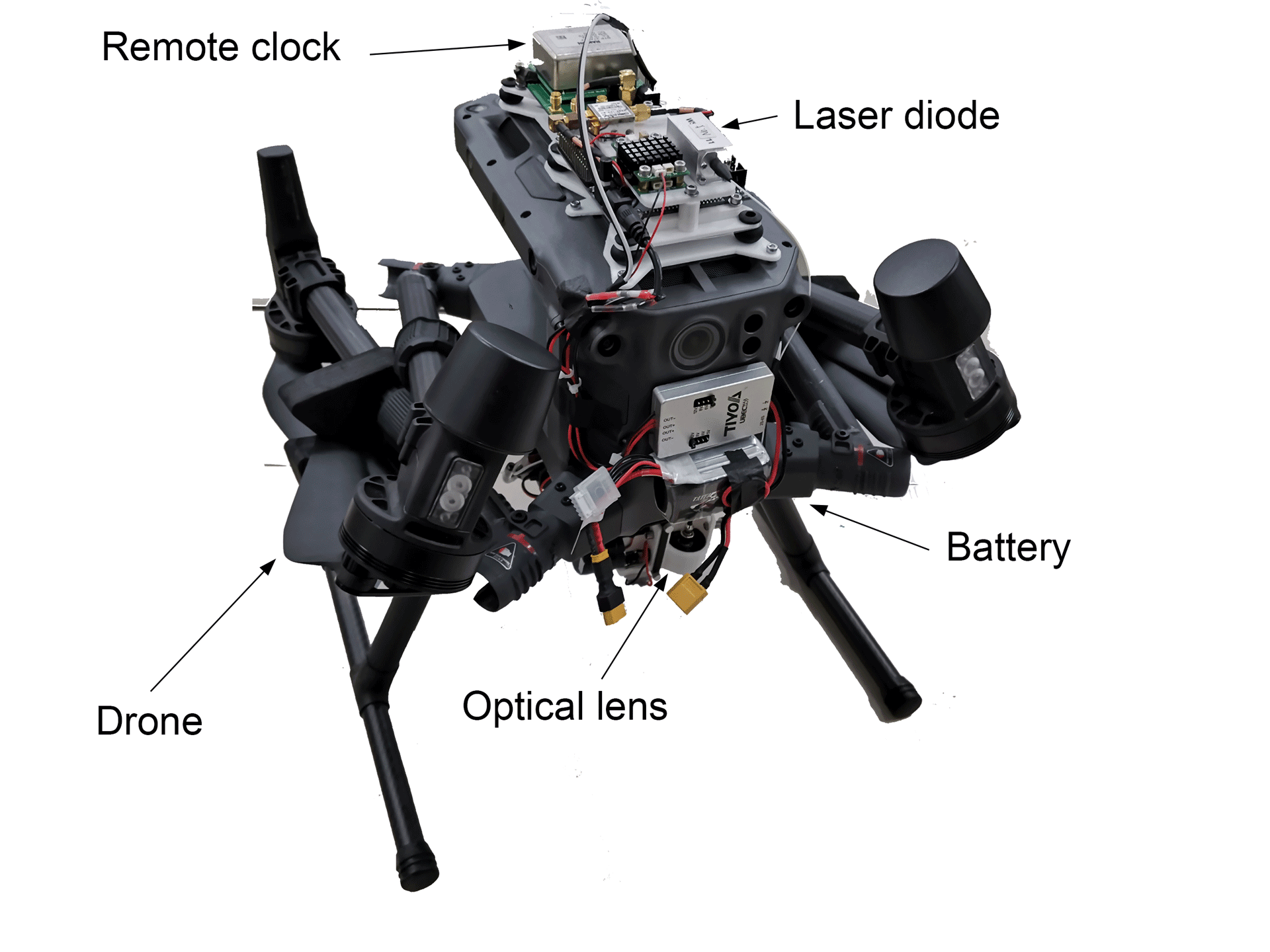}
    \caption{\textbf{Onboard transmitter of the outdoor verification experiment.}
    The PICOC transmitter consists of a common remote clock, a laser diode, a battery, and a set of optical lenses.
    }
    \label{f1}
\end{figure*}

By setting the drone's height to 500m, we expect that the light spot diameter expands to about 1000m (see Fig.S3b). We use a 2cm-aperture lens to couple photons into a multi-mode optical fiber and detect photons using a single-photon detector at the receiving terminal. Background noise, no matter from stars, street lamps, or spotlights, is around 3500 photons per second (see Fig.S3d). By adjusting the course angle and the horizontal position of the drone, we collect data with relatively low signal intensity. The sky-blue surface in Fig.S3a shows the theoretical time evolution of photon statistics, and dark blue curves show the experimental results. The experimental results justify our assumption well about the properties of the measurement process (see Fig.S3a and Fig.S3c).

\begin{figure*}[htbp]
    \centering
    \includegraphics[width=1.0\linewidth]{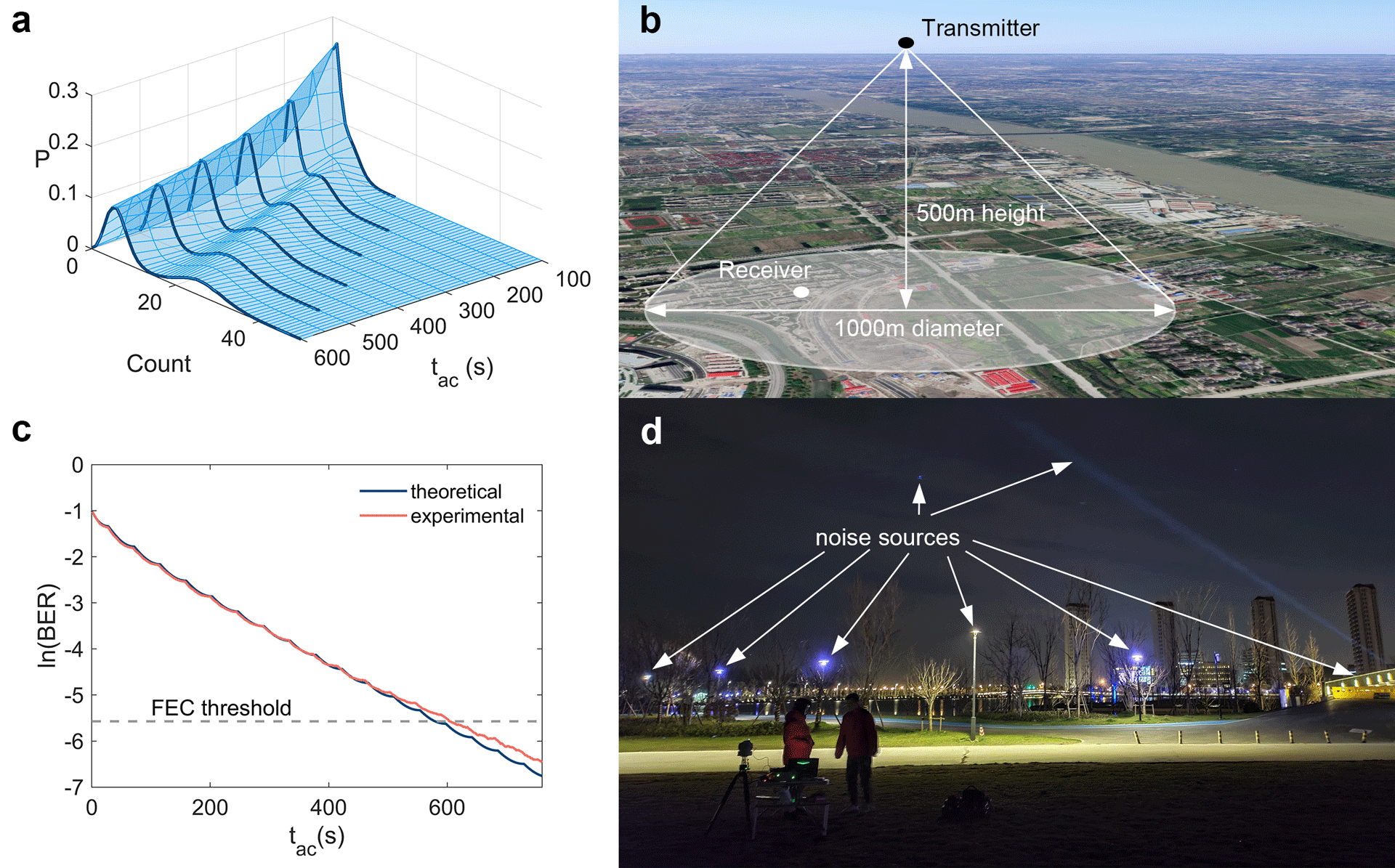}
    \caption{\textbf{Outdoor verification on the properties of the measurement process.}
    \textbf{ a}, Time evolution of photon statistics.
    \textbf{ b}, Illustration of the verification experiment.
    \textbf{ c}, Theoretical and experimental results of the time evolution of bit error rate.
    \textbf{ d}, Noise sources in the verification experiment. BER: bit error rate.
    }
    \label{f1}
\end{figure*}

\subsection*{3. Channel loss in the PICOC system}
In the interstellar communication scenario, the geometric loss contributes most of the channel loss. However, atmospheric absorption and turbulence are also essential factors for interstellar communication. Above the 100km height, the air density is lower than the air density at ground level. Absorption and turbulence loss caused by the upper atmosphere are negligible. We expect the loss resulting from the atmosphere absorption and turbulence is around 3dB\cite{QKD_satellite}. The channel loss can be expressed as
\begin{equation}
    \left\{\begin{array}{l}
    Loss=10 \log \left[\eta\left(1-e^{-\frac{D^{2}}{2 \omega_{z}^{2}}}\right)\right] \\
    \omega_{z}=\omega_{0} \sqrt{1+\frac{\lambda z}{\pi \omega_{0}^{2}}}
    \end{array}\right.
\end{equation}
where $\eta$ represents the atmosphere transmittance considering atmosphere absorption and turbulence. $D$ represents the diameter of the receiving telescope. $\omega_{0}$ represents the beam waist of the transmitting telescope. $\lambda$ is the wavelength of the light. $z$ is the distance between the transmitter and the receiver.

\begin{figure*}[!h]
    \centering
    \includegraphics[width=1.0\linewidth]{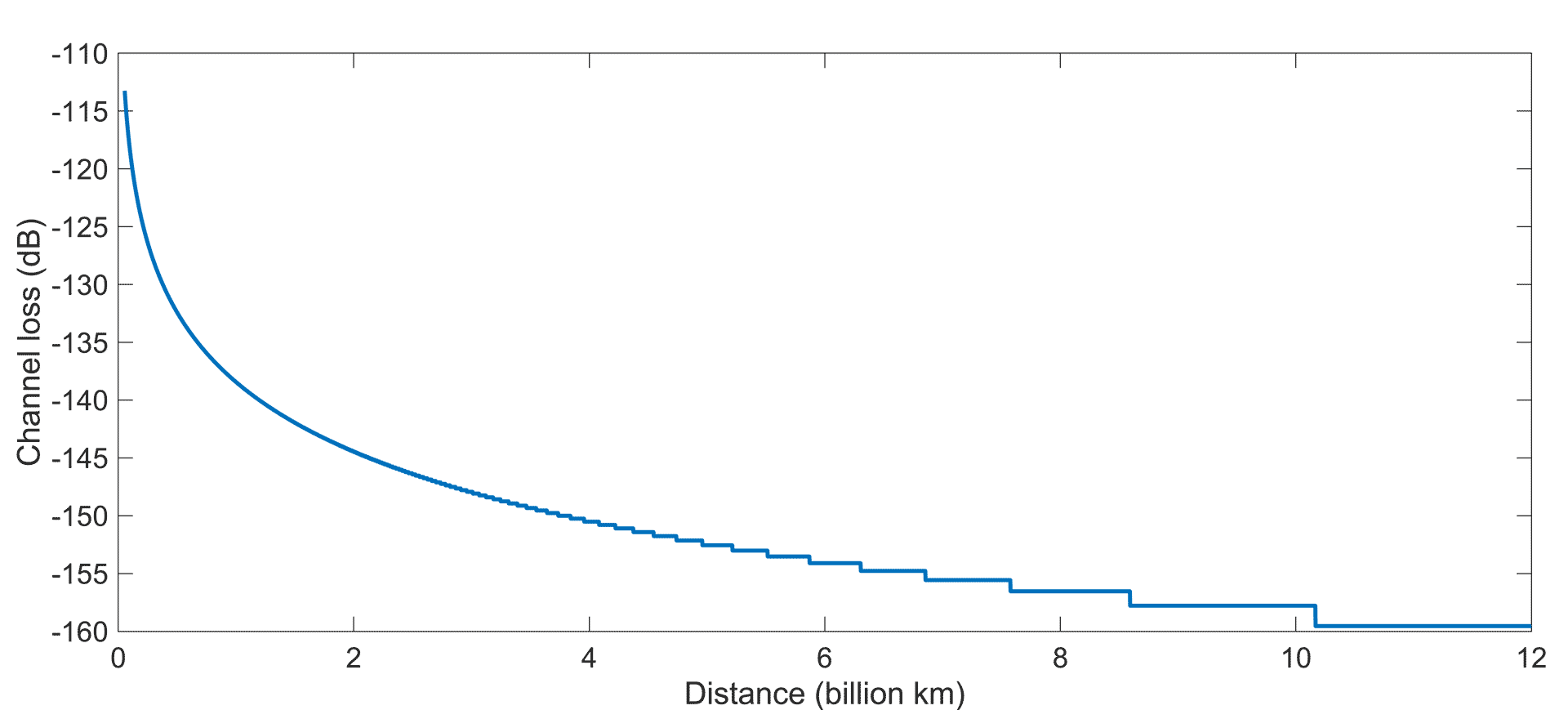}
    \caption{\textbf{The relationship between channel loss and communication distance in PICOC system.}
    We show the corresponding channel loss from the 55 million kilometers to 12 billion kilometers.
    }
    \label{f1}
\end{figure*}

We take the transmitting telescope of the satellite Quantum Experiments at Space Scale as reference\cite{QKD_satellite} and set $\omega_{0}=0.054m$, $\lambda=850nm$. We use a 1.2m aperture telescope in the receiving terminal to collect photons and set $D=1.2m$. $\eta=0.5$ represents the 3dB loss caused by atmosphere absorption and turbulence\cite{QKD_satellite}. Fig.S4 shows the relationship between channel loss and communication distance. The channel loss would achieve 160dB when the communication distance reaches 12 billion kilometers.

\subsection*{4. Time synchronization of PICOC under low signal gain}
We show the robustness of PICOC's time synchronization scheme against the low signal gain and noisy environments. Fig.S5 to Fig.S8 show the retrieval of time frames from the sparse single-photon stream, with only $5.4 \times 10^{-5}$ signal photon per pulse is received. When the noise level is extremely low, we can observe an apparent periodic oscillation of coincidence delay over time (see Fig.S5a). By correlating the overall coincidence delay, we manage to precisely characterize the linear phase drift between remote clocks (see Fig.S5b).

\begin{figure*}[!h]
    \centering
    \includegraphics[width=1.0\linewidth]{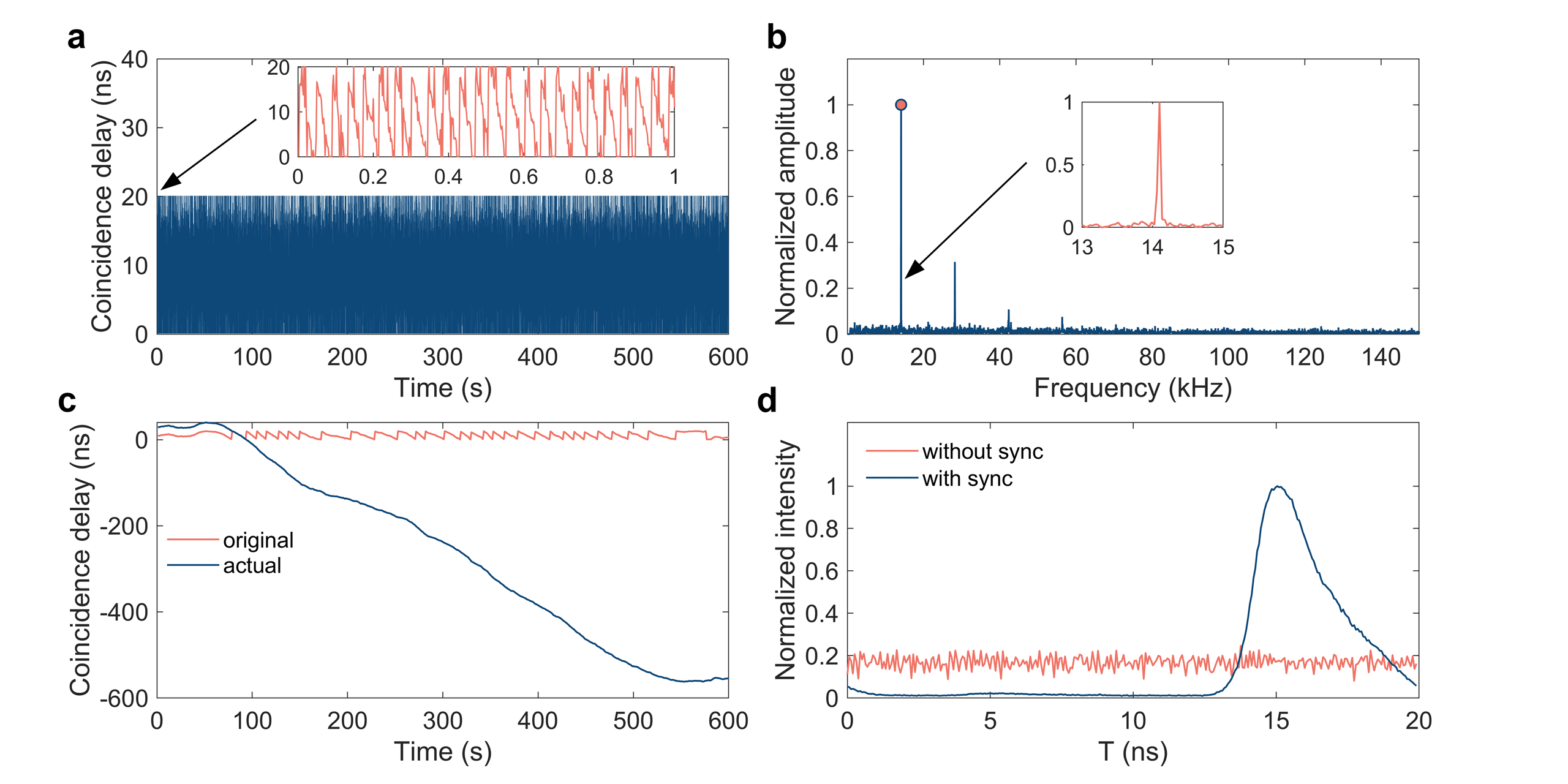}
    \caption{\textbf{The time synchronization performance of PICOC under low signal gain and low noise.}
    \textbf{ a}, Laser output power 0.552mW; modulation rate 50MHz; signal intensity is around 1000 photons per second; noise intensity is around 70 photons per second. Stage1: We retrieve the correlation shared among a few photons and search the coincidence delay of each time slot.
    \textbf{ b}, Stage2: We characterize the global linear phase drift between remote clocks by performing a Fourier transformation.
    \textbf{ c}, Stage3: We characterize the global non-linear phase drift between remote clocks.
    \textbf{ d}, We retrieve the waveform of the pulse with and without our scheme.
    }
    \label{f1}
\end{figure*}
It should be noted that only the first peak (rather than the highest peak) in Fig.5b represents the practical linear phase drift. The second peak that characterizing the double frequency is not an actual oscillation mode.

With the rise of noise level, time frames share among survived signal photons are affected by noise photons. The oscillation of coincidence delay over time becomes unclear (see Fig.S6a). See also the incomplete characteristic peaks shown in Fig.S6b. However, we can still retrieve the high-precision time synchronization over the 800s duration measurement (see Fig.S6d).

\begin{figure*}[!h]
    \centering
    \includegraphics[width=1.0\linewidth]{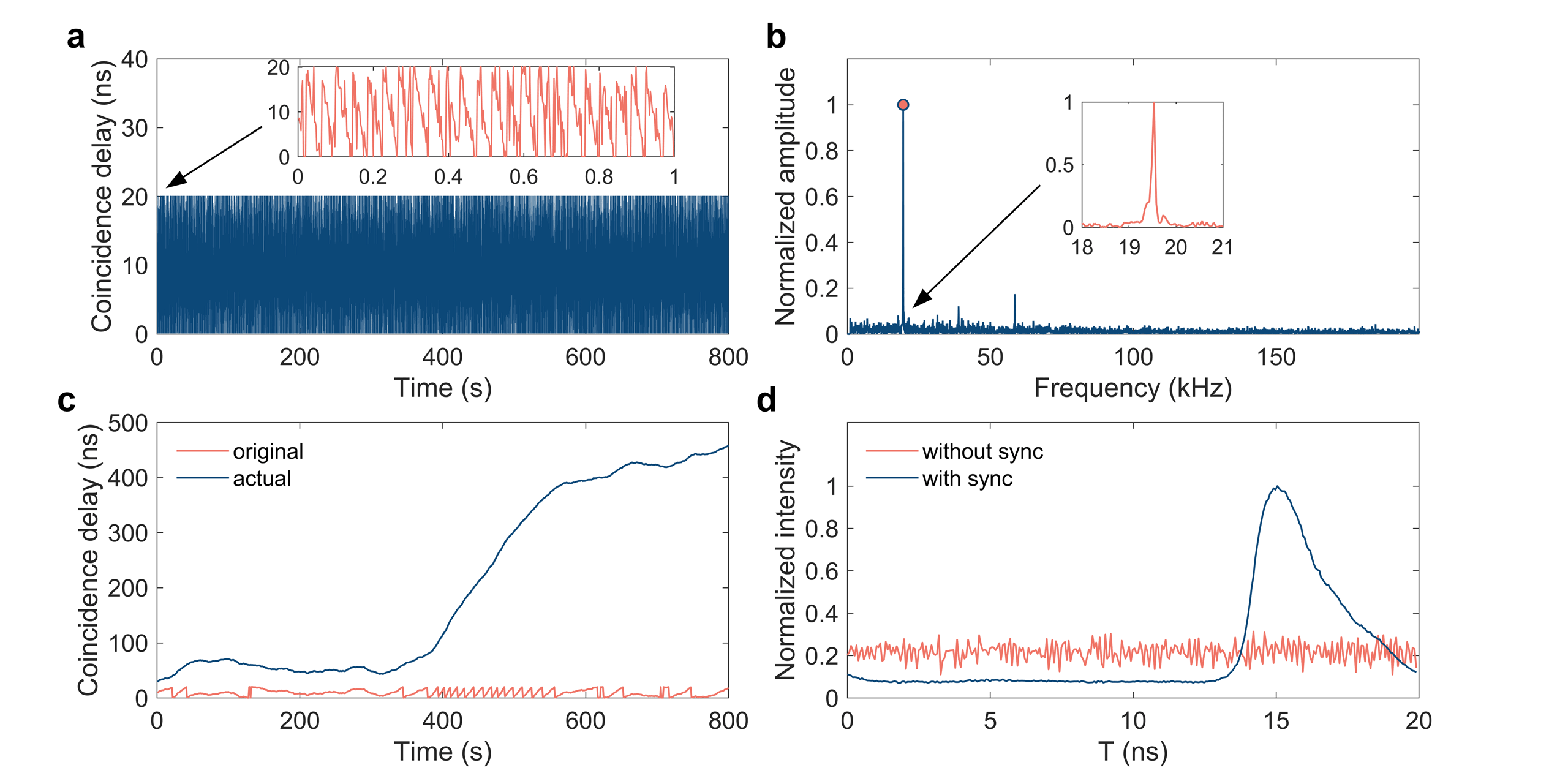}
    \caption{\textbf{The time synchronization performance of PICOC under low signal gain and middle noise.}
    \textbf{ a}, Laser output power 0.552mW; modulation rate 50MHz; signal intensity is around 1000 photons per second; noise intensity is around 580 photons per second. Stage1: We retrieve the correlation shared among a few photons and search the coincidence delay of each time slot.
    \textbf{ b}, Stage2: We characterize the global linear phase drift between remote clocks by performing a Fourier transformation.
    \textbf{ c}, Stage3: We characterize the global non-linear phase drift between remote clocks.
    \textbf{ d}, We retrieve the waveform of the pulse with and without our scheme.
    }
    \label{f1}
\end{figure*}

When the signal-to-noise ratio (SNR) drops below 0dB, noise photons contribute more information than signal photons. We can hardly recognize the phase drift between remote clocks with the naked eye (see Fig.S7a). Simultaneously, our PICOC time synchronization scheme still works fine (see Fig.S7b and Fig.S7c). The retrieved pulse width is 6ns, close to the transmitted pulse width (see Fig.S7d).

\begin{figure*}[!h]
    \centering
    \includegraphics[width=1.0\linewidth]{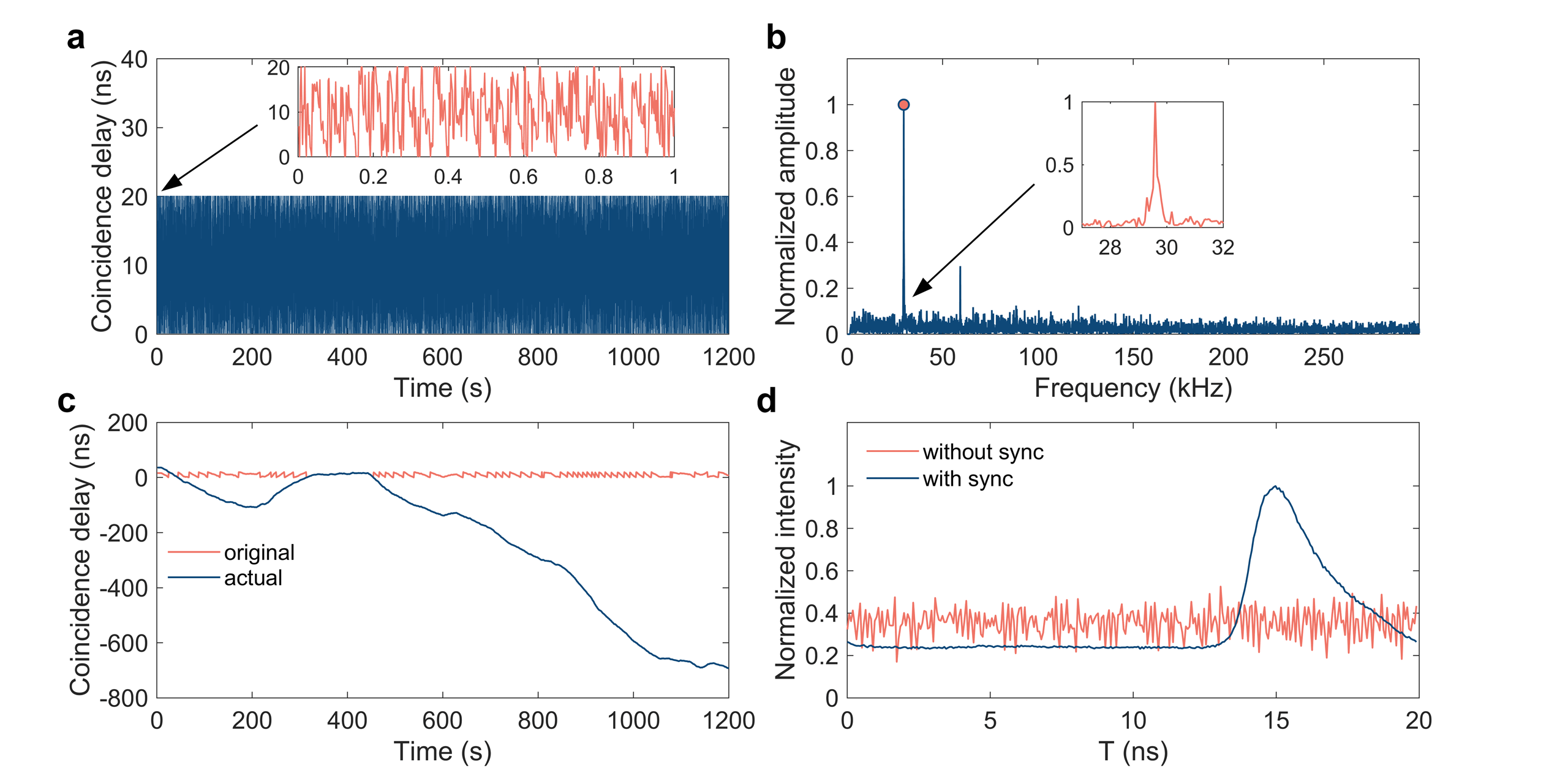}
    \caption{\textbf{The time synchronization performance of PICOC under low signal gain and high noise.}
    \textbf{ a}, Laser output power 0.552mW; modulation rate 50MHz; signal intensity is around 1000 photons per second; noise intensity is around 2000 photons per second. Stage1: We retrieve the correlation shared among a few photons and search the coincidence delay of each time slot.
    \textbf{ b}, Stage2: We characterize the global linear phase drift between remote clocks by performing a Fourier transformation.
    \textbf{ c}, Stage3: We characterize the global non-linear phase drift between remote clocks.
    \textbf{ d}, We retrieve the waveform of the pulse with and without our scheme.
    }
    \label{f1}
\end{figure*}

By introducing more noise photons, we demonstrate the retrieval of time frames under the -10.35dB SNR environment. Noise photons contribute most of the information, letting us incapable of observing the oscillation feature under this circumstance (see Fig.S8a). However, the global correlation feature of PICOC allows us to analyze the internal correlation shared for a very long time. The characteristic peak shown in Fig.S8b is almost drowned in noise, but we can still calibrate the linear phase drift correctly. By further calibrate the long-term non-linear phase drift (see Fig.S8c), we show the strong robustness of our scheme against extremely low SNR conditions (see Fig.S8d).

\begin{figure*}[!h]
    \centering
    \includegraphics[width=1.0\linewidth]{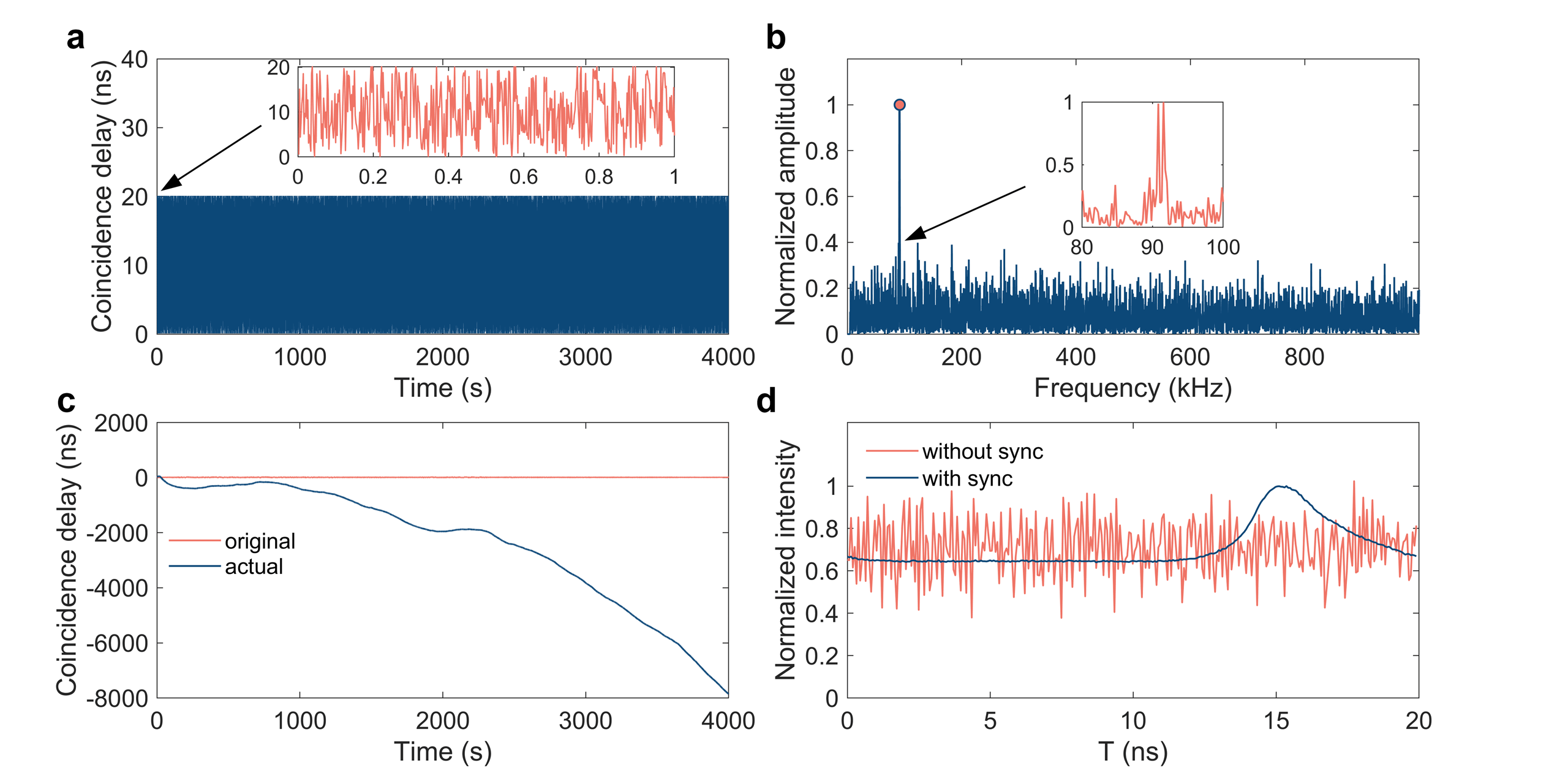}
    \caption{\textbf{The time synchronization performance of PICOC under low signal gain and ultra-high noise.}
    \textbf{ a}, Laser output power 0.552mW; modulation rate 50MHz; signal intensity is around 1000 photons per second; noise intensity is around 10000 photons per second. Stage1: We retrieve the correlation shared among a few photons and search the coincidence delay of each time slot.
    \textbf{ b}, Stage2: We characterize the global linear phase drift between remote clocks by performing a Fourier transformation.
    \textbf{ c}, Stage3: We characterize the global non-linear phase drift between remote clocks.
    \textbf{ d}, We retrieve the waveform of the pulse with and without our scheme.
    }
    \label{f1}
\end{figure*}

In Fig.S9, we show the time synchronization process under ultra-low signal gain, with only $1.5 \times 10^{-5}$ signal photon per pulse is received. 2ms duration time slots divide a second evenly into 500 pieces. Owing to the ultra-low signal gain, only half of the time slot can characterize the coincidence delay. As is shown in Fig.S9a, errors caused by empty time slots cover the oscillation feature of coincidence delay over time. By correlating the long-term internal time information, we manage to find the characteristic peak and calibrate the linear phase drift (see Fig.S9b). From the calibrated non-linear phase drift curve shown in Fig.S9c, we can observe an obvious linear correlation between the coincidence delay and time. This means that the linear calibration part is not performing well. However, the phase drift rate shown in Fig.S9c is relatively tiny compared with the phase drift rate shown in Fig.S9b. We can still synchronization the extremely weak signal with high precision (see Fig.S9d).

\begin{figure*}[!h]
    \centering
    \includegraphics[width=1.0\linewidth]{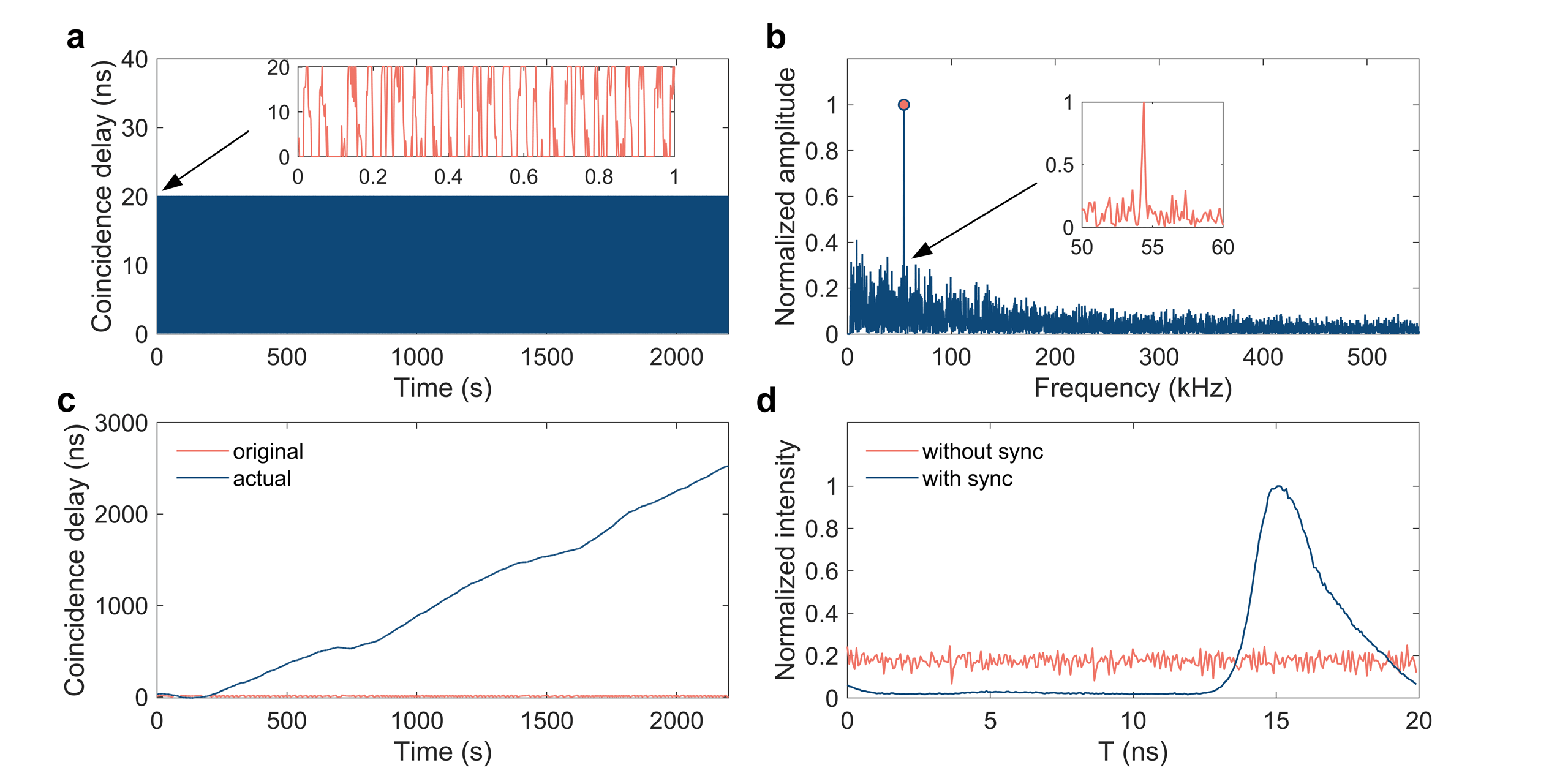}
    \caption{\textbf{The time synchronization performance of PICOC under ultra-low signal gain and low noise.}
    \textbf{ a}, Laser output power 0.552mW; modulation rate 50MHz; signal intensity is around 280 photons per second; noise intensity is around 30 photons per second. Stage1: We retrieve the correlation shared among a few photons and search the coincidence delay of each time slot.
    \textbf{ b}, Stage2: We characterize the global linear phase drift between remote clocks by performing a Fourier transformation.
    \textbf{ c}, Stage3: We characterize the global non-linear phase drift between remote clocks.
    \textbf{ d}, We retrieve the waveform of the pulse with and without our scheme.
    }
    \label{f1}
\end{figure*}

\subsection*{5. The 0dB SNR boundary}
The 0dB SNR is very common in traditional optical communication. However, when the channel loss is extremely high and most of the signal pulses are lost, the 0dB SNR should be treated as an interesting boundary.

In traditional optical communication, we usually use PIN photodiode to detector signals. The outputs of a PIN detector are voltage output. By detecting the rising edge of pulses, we can determine empty pulses and non-zero pulses (see Fig.S10a). In this case, the tolerable SNR can be much lower than -10dB, as long as the detector can respond to the pulse rising edge correctly. However, optical communication systems using photodiodes as detectors can only establish links against low channel loss, restricting the application of long-distance free-space optical communication.

With the increase of channel loss, the received signal intensity drops. When the signal intensity drops to the dozens of photons per pulse, the single-photon detector (SPD) is introduced in detecting photons. The outputs of the SPD are narrow electrical pulses, and each pulse represents a single-photon event. In this case, the 0dB SNR becomes an interesting threshold because noise photons and signal photons are identical (see Fig.S10b). We cannot tell a signal photon from the noise without introducing an extra degree of freedom of light. As far as we know, establishing reliable single-photon level optical communication links against 0dB SNR is still beyond reach.

In our case, the tolerable channel loss is extremely high and most of the signal pulses are lost. The received signal intensity drops to $10^{-5}$ photon per pulse. Only about one single-photon event locates in $10^{5}$ pulses on average (see Fig.S10c). The 0dB and -10dB SNR represent that the possibility of a single-photon event being a signal photon is 50$\%$ and 10$\%$ respectively. We cannot distinguish signal from noise only by counting the photon number in each pulse. However, the global correlation feature and the dynamic photon statistics behavior of PICOC allow us to retrieve signals embedded in noise ten times higher and establish optical communication links against 160dB channel loss and -10dB SNR.

\begin{figure*}[!h]
    \centering
    \includegraphics[width=1.0\linewidth]{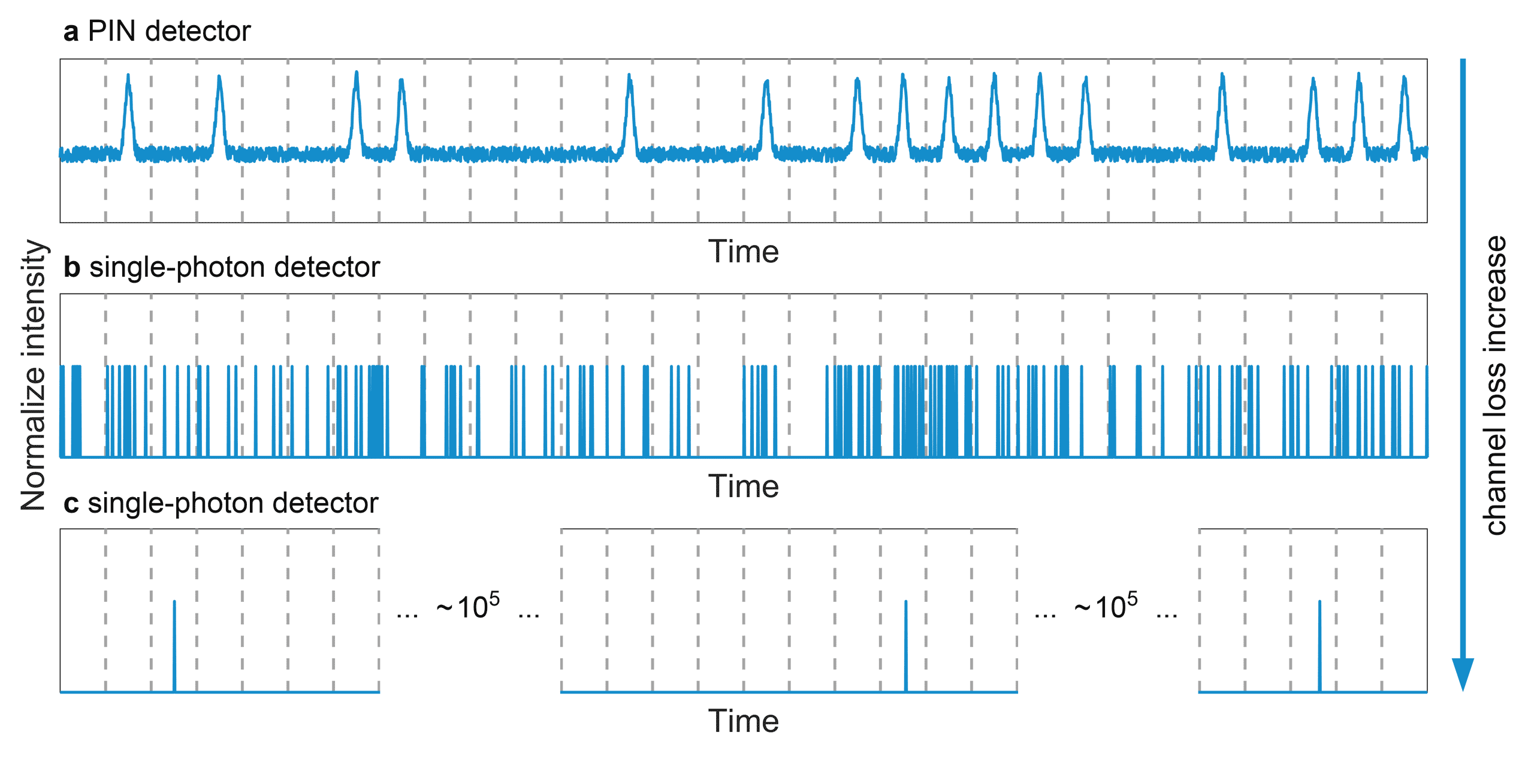}
    \caption{\textbf{Illustration of the received signal using different detectors.}
    \textbf{ a}, The outputs of a PIN photodiode with extremely high signal gain.
    \textbf{ b}, The outputs of a single-photon detector when the signal intensity is around dozens of photons per pulse.
    \textbf{ c}, The outputs of a single-photon detector when the received signal is extremely sparse.
    }
    \label{f1}
\end{figure*}

\subsection*{6. Derivation of bit error rate in PICOC system}
\subsubsection*{Photon statistics of PICOC bits}
The photon number of pulses at receiving terminal follows a Poisson distribution with a parameter $\mu_{sp}^{\prime}$\cite{XiongfengMa:QKD,Hoi_Kwong_Lo:QKD}. Therefore, the photon statistics can be expressed as
    \begin{equation}
        P_{s p}\left(\mu_{s p}^{\prime}, n\right)=\frac{e^{-\mu_{s p}^{\prime}} \mu_{s p}^{\prime n}}{n !}
    \end{equation}
where $n$ indicates the photon number.

Define $\eta$ as the detection efficiency of SPD. The SPD can be modeled as a threshold detector, which means the detector can tell the difference of zero photon and non-zero photon, instead of the actual received photon number\cite{XiongfengMa:QKD}. Thus, the detection probability of n-photon state is given by
    \begin{equation}
        \eta_{n}=1-(1-\eta)^{n}(n=0,1,2,\cdots)
    \end{equation}
The typical value of $\mu_{sp}^{\prime}$ is $\sim10^{-5}$ photon per pulse. The expected value of detection events is given by
    \begin{equation}
        \begin{aligned}
            \mu_{s p} &=\sum_{n=0}^{\infty}\left(\frac{e^{-\mu_{s p}^{\prime}} \mu_{s p}^{\prime n}}{n !} \cdot n \cdot \eta_{n}\right) \\
            & \approx \mu_{s p}^{\prime} \eta
        \end{aligned}
    \end{equation}
Therefore, the statistics of signal pulses follows a Poisson distribution with a parameter $\mu_{sp}$.
    \begin{equation}
        P_{s p}\left(\mu_{s p}, n\right)=\frac{e^{-\mu_{s p}} \mu_{s p}}{n !}
    \end{equation}

Owing to the small value of $\mu_{sp}$, photon-inter-correlation scheme is applied to ensure enough signal events detected for each bit. Therefore, statistics of a 1-bit per second is given by
    \begin{equation}
        P\left(\mu_{s p}, m, n\right)=\frac{1}{n !}\left(m \mu_{s p}\right)^{n} e^{-m \mu_{s p}}
    \end{equation}
where $m$ represents the correlated pulse number of a 1-bit during the time span of one second. The statistics of 1-bits is given by

    \begin{equation}
        P\left(\mu_{s p}, m, t_{ac}, n\right)=\frac{e^{-\mu_{sp} m t_{ac}}\left(\mu_{s p} m t_{ac}\right)^{n}}{n !}
    \end{equation}
Define $\mu_{s}=m\mu_{sp}$, then
    \begin{equation}
        P\left(\mu_{s}, t_{ac}, n\right)=\frac{e^{-\mu_{s} t_{ac}}\left(\mu_{s} t_{ac}\right)^{n}}{n !}
    \end{equation}
The statistics still follows a Poisson distribution. By extending the accumulation time $t_{ac}$, ultra-high channel loss can be tolerable. Meanwhile, signal photons completely embedded in noise can also be retrieved.

The noise mainly comes from the detector and environment background. It is reasonable to assume that the noise is random in time span and the statistics of noise can be expressed as
    \begin{equation}
        P\left(\mu_{n}, t_{ac}, n\right)=\frac{e^{-\mu_{n} t_{ac}}\left(\mu_{n} t_{ac}\right)^{n}}{n !}
    \end{equation}
The noise contributes to both 1-bits and 0-bits, while the detection events of true signal only affects 1-bits. Thus, the statistics of 1-bits and 0-bits can be expressed as
    \begin{equation}
        \left\{
        \begin{aligned}
            P_{1}\left(t_{ac}, n\right)&=\frac{e^{-\mu_{1} t_{ac}}\left(\mu_{1} t_{ac}\right)^{n}}{n !} \\
            P_{0}\left(t_{ac}, n\right)&=\frac{e^{-\mu_{0} t_{ac}}\left(\mu_{0} t_{ac}\right)^{n}}{n !} \\
            \mu_{1}&=\mu_{s}+\mu_{n} \\
            \mu_{0}&=\mu_{n}
        \end{aligned}
        \right.
    \end{equation}

\subsubsection*{The optimization of $N_{s}$}
BER of PICOC is given by
    \begin{equation}
        BER(t_{ac})=a_{0}\sum\limits_{n=N_{s}}^{\infty}\frac{(\mu_{0} t_{ac})^{n}}{n!}e^{-\mu_{0} t_{ac}}+a_{1}\sum\limits_{n=0}^{N_{s}-1}\frac{(\mu_{1} t_{ac})^{n}}{n!}e^{-\mu_{1} t_{ac}}
    \end{equation}
The optimal $N_{s}$ minimizing the BER is given by
    \begin{equation}
        N_{s}(t_{ac})=\left\lceil\frac{\ln \left(a_{0} / a_{1}\right)+t_{ac}\left(\mu_{1}-\mu_{0}\right)}{\ln \left(\mu_{1} / \mu_{0}\right)}\right\rceil
    \end{equation}
From which we can observe the BER decreases periodically with $t_{ac}$. The oscillation period and phase of BER is expressed as
    \begin{equation}
        \left\{\begin{aligned}
            T_{N_{s}}=\frac{\ln \left(\mu_{1} / \mu_{0}\right)}{\mu_{1}-\mu_{0}} \\
            \varphi_{N_{s}}=\frac{\ln \left(a_{0} / a_{1}\right)}{\mu_{0}-\mu_{1}}
        \end{aligned}
        \right.
    \end{equation}

\subsubsection*{The approximated BER expression of PICOC}
Approximately, the BER of PICOC can be expressed as
    \begin{equation}
        \left\{\begin{aligned}
            \ln \left[BER(t_{ac})\right]&=G_{cc} \left( t_{ac} +\varphi_{N_{s}} \right) +p+A_{p}\left|\sin \left[\omega\left(t_{a c}-\varphi_{N_{s}}\right)\right]\right| \\
            G_{cc}&=\delta_{j} / T_{N_{s}} \\
            w&=\pi / T_{N_{s}} \\
            p&=\ln \left( \frac{a_{1}}{2}\right) / 2 + \Delta_{j} \\
            A_{p}&=A_{j}/2
        \end{aligned}\right.
    \end{equation}
The above expression consists of two terms, the linear term and the oscillating term. The linear term treats light as a continuous field, and the oscillating term takes the indivisibility of single photons into account.

Here, we set $t_{j}=\varphi_{N_{s}}+j T_{N_{s}}$ ($j=0,1,2,\cdots$ represents the $j^{th}$ oscillation period) and $N_{s,j}=\left\lceil j \right\rceil = j$. Assume that the ratio of 0-bits and 1-bits are $a_{1}=a_{0}=0.5$ respectively. The corresponding parameters are defined as
    \begin{equation}
        \left\{\begin{aligned}
            \delta_{j} &=\ln \left(E_{t_{j+1}}\right) - \ln \left(E_{t_{j}}\right) \\
            \Delta_{j} &=\ln \left(E_{t_{j+1}}\right)-\ln \left(E_{t_{0}}\right)-(j+1) \delta_{j} \\
            A_{j} &=\ln \left(E_{t_{j+1/2}}\right) - \ln \left(E_{t_{j}}\right) \\
        \end{aligned}
        \right.
    \end{equation}
The curve of BER experiences more oscillation periods under lower SNR environments and vice versa. Normally $j=20$ is great enough for the low SNR environment, while $j=5$ is suitable for the high SNR environment.

\clearpage

\end{document}